\documentclass[3p,times,twocolumn]{elsarticle}

\usepackage{ecrc}

\volume{00}

\firstpage{1}

\journalname{Nuclear and Particle Physics Proceedings}

\runauth{}

\jid{nppp}

\jnltitlelogo{Nuclear and Particle Physics Proceedings}

\usepackage{amssymb}
\usepackage{amsmath}
\usepackage{slashed}
\usepackage{xspace}
\usepackage{amsthm}

\usepackage[figuresright]{rotating}

\def\mpt{\slashed{p}^\parallel_{T}}
\def\Nob{\Psi^{S}_{P_T} (r)}

\begin{document}

\begin{frontmatter}



\dochead{}

\title{The angular structure of jet quenching within a hybrid strong/weak coupling model}


\author[a]{Jorge Casalderrey-Solana}
\author[c]{Doga Can Gulhan}
\author[e,f]{Jos\'e Guilherme Milhano}
\author[b]{Daniel Pablos}
\author[g]{Krishna Rajagopal}

\address[a]{
Rudolf Peierls Centre for Theoretical Physics, University of Oxford, 1 Keble Road, Oxford OX1 3NP, United Kingdom}

\address[b]{
Departament de F\'\i sica Qu\`antica i Astrof\'\i sica \&  Institut de Ci\`encies del Cosmos (ICC), Universitat de Barcelona, Mart\'{\i}  i Franqu\`es 1, 08028 Barcelona, Spain}

\address[c]{CERN, EP Department, CH-1211 Geneva 23, Switzerland}

\address[e]{Laborat\'orio de Instrumenta\c c\~ao e F\'isica Experimental de Part\'iculas (LIP), Av. Elias Garcia 14-1, P-1000-149 Lisboa, Portugal}
\address[f]{Theoretical Physics Department, CERN, Geneva, Switzerland}

\address[g]{Center for Theoretical Physics, Massachusetts Institute of Technology, Cambridge, MA 02139 USA}

\begin{abstract}
Building upon the hybrid strong/weak coupling model for jet quenching, we incorporate and study the effects of transverse momentum broadening and medium response of the plasma to  jets on a variety of observables. 
For inclusive jet observables, we find little sensitivity to the strength of broadening. To constrain those dynamics, we propose new observables constructed from ratios of differential jet shapes, in which particles are binned in momentum, which are sensitive to the in-medium broadening parameter. We also investigate the effect of the back-reaction of the medium on the angular structure of jets as reconstructed with different cone radii $R$. Finally we provide results for the so called ``missing-pt'', finding a qualitative agreement between our model calculations and data in many respects, although a quantitative agreement is beyond our simplified treatment of the  hadrons originating from the hydrodynamic wake.
\end{abstract}


\end{frontmatter}


\subsection*{Introduction}
\label{intro}

The discovery that the quark-gluon plasma (QGP) created in heavy ion collisions behaves as a strongly coupled fluid has opened a rich window of phenomenology for holographic techniques. One of the most useful tools to study the properties of the QGP is given by the analysis of the modification of high energy jets that traverse the exploding fireball. The evolution of these energetic excitations within medium is governed by a wide range of scales, from the hard virtuality of the perturbative production and consequent branching, to the relatively low temperature of the plasma that interacts strongly with the jet.\

Based on this observation we have developed a model for jet quenching that incorporates aspects of weak and strong coupling physics applied at the scale where they are expected to be valid \cite{Casalderrey-Solana:2014bpa,Casalderrey-Solana:2015vaa,Casalderrey-Solana:2016jvj}. The core assumptions around which the construction of the model is performed are basically two. First, the evolution of the parton shower follows the DGLAP equations due to the decoupling of virtuality and temperature scales, and second, between splittings the partons interact strongly with the QGP by transferring energy and momentum to it as dictated by semiclassical string calculations derived in gauge/gravity duality \cite{Chesler:2014jva,Chesler:2015nqz}.\

The simple and phenomenological approach taken in this one-parameter hybrid model has allowed for a systematic comparison against experimental data (see for instance photon-jet comparison in \cite{CMSphoton}), providing insights into the relevance of the different physical mechanisms that potentially play a role in the jet/plasma interplay. We present next an analysis of the effects that transverse momentum broadening and medium response to energy and momentum deposition induce on a selection of jet and intra-jet observables.

\begin{figure}[t]
\centering 
\includegraphics[width=.5\textwidth]{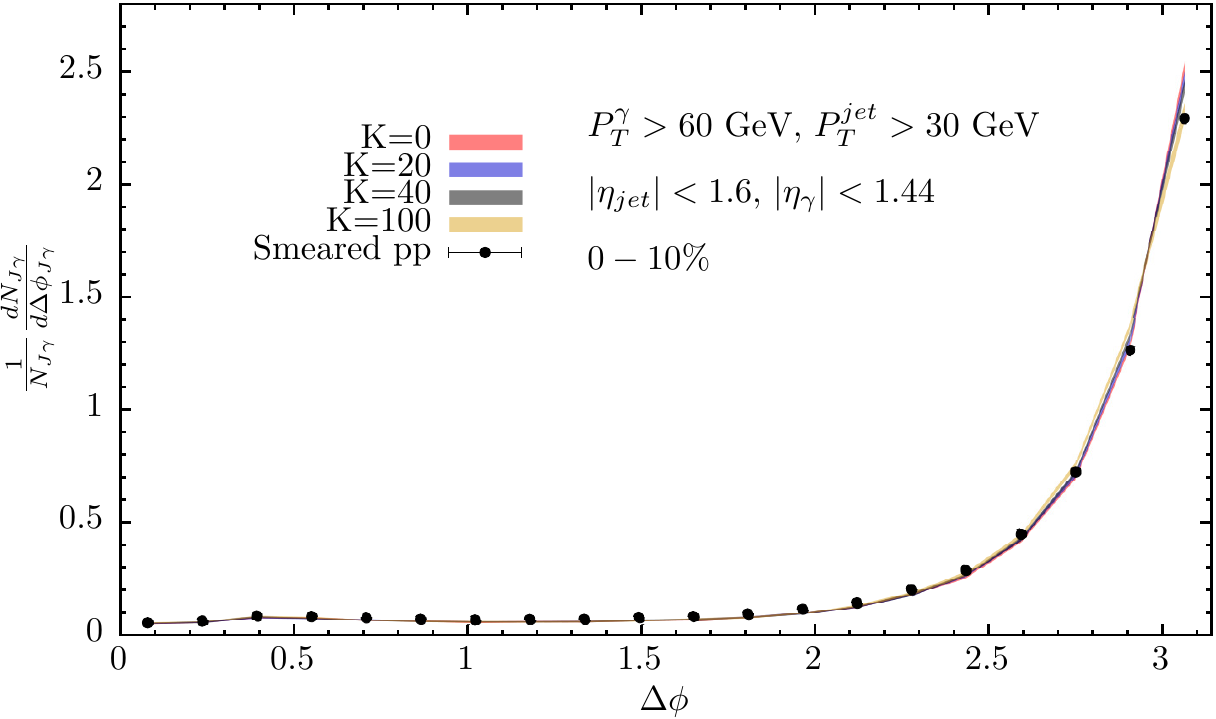}
\caption{$\Delta\phi$ distributions in the transverse plane between an isolated photon and an anti-$k_t$ jet with radius $R=0.3$ satisfying the cuts, both for vacuum (smeared pp) and for the $0-10\%$ centrality class PbPb collisions. Different values of the broadening parameter $K$ are explored.}
\label{fig1}
\end{figure}

\subsection*{Transverse momentum broadening}
\label{broad}
In the strongly coupled limit there is no notion of scattering centers and no notion
of multiple discrete transfers of momentum - the perturbative scheme under which a parton traversing a hot medium would experience transverse momentum broadening according to a Gaussian distribution. However, it has been shown in \cite{Liu:2006ug,DEramo:2010wup} within the large 
't~Hooft coupling limit $\lambda$, that coloured excitations acquire transverse momentum following also a Gaussian distribution with a width $Q^2_\perp= \hat q L$, where $\hat q \propto \sqrt{\lambda}T^3$. Given that at strong coupling there is no strong correlation between the dynamics generating transverse momentum broadening and that responsible for energy loss, we add to our model the free parameter $K$ to gauge the width of the Gaussian distribution through $\hat q = K T^3 $. Perturbative analyses of jet quenching have quoted a value of this parameter of roughly $K\sim 5$, while from strong coupling analyses one would expect $K \sim 20$ (more details in \cite{Casalderrey-Solana:2016jvj}).\

Rather surprisingly at first, the sensitivity of standard observables to the inclusion of broadening mechanism (even for the extreme case $K=100$) is fairly small. In Fig. \ref{fig1} we show an example of such insensitivity captured by the photon-jet acoplanarity (the distribution in angular separation $\Delta\phi$ in the transverse plane between an isolated photon and the rest of the jets in the event which satisfy the momentum and pseudorapidity cuts). The already wide vacuum distribution, which differs from a $\delta$ function at $\Delta\phi=\pi$ due to initial state radiation, final state radiation and non-prompt photon contamination (besides experimental smearing effects), is barely modified by the medium contribution coming from the convolution with the transverse kicks Gaussian distribution. Actually, what one observes is a slight narrowing of the medium distributions as a consequence of energy loss, which results in a greater suppression of wider jets with more partonic activity, which are the jets that are more deflected, relative to narrower, less acoplanar, jets.\

\begin{figure}[t]
\centering 
\includegraphics[width=.5\textwidth]{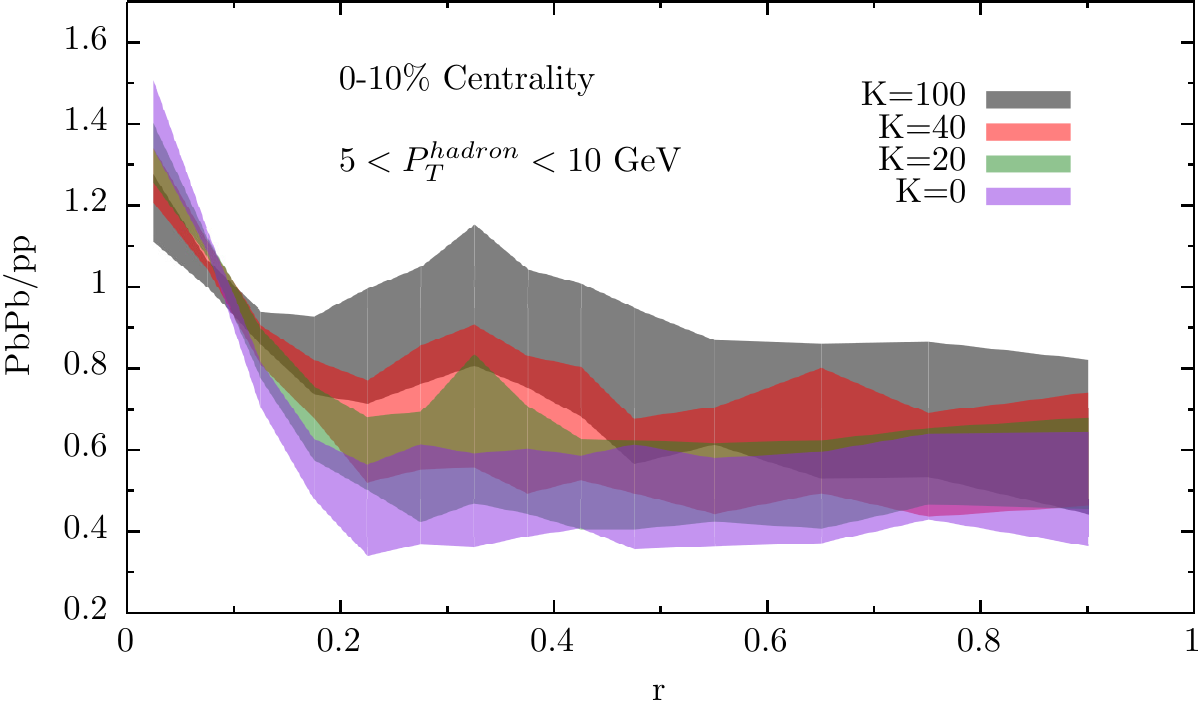}
\caption{The ratio of PbPb over pp for the special jet shapes $\Nob$ for $0-10\%$ centrality class. The jet sample is made of $R=0.3$ subleading jets in a dijet pair, and the hadrons entering the histogram are restricted to be within the range $5<P_T<10$ GeV in order to focus the observable on those tracks for which the effects of broadening are large.}
\label{fig2}
\end{figure}

To study in-medium broadening, we have designed a special observable, that we shall call $\Nob$, which shows a remarkable sensitivity to the broadening parameter $K$. It is shown in Fig. \ref{fig2}, and corresponds to a modified version of the PbPb over pp ratio of standard jet shapes, which quantifies the medium modification of the relative contribution to the total jet energy as a function of the angular separation $r$ in the $(\phi,\eta)$ plane with respect to the jet axis. The jets entering the plot shown in Fig. \ref{fig2} are the subleading jets in a dijet pair (with $P_T^L>120$ GeV, $P_T^S>30$ GeV and $\Delta\phi>5 \pi/6$) reconstructed using anti-$k_t$ with jet radius $R=0.3$. While the total energy of the jet is reconstructed using all tracks, we restrict the momentum range of the particles that enter the $r$ distribution to be between $5<P_T<10$ GeV. The precise expression for this observable is
\begin{equation}
\Nob\equiv \frac{1}{N_{\textrm{S}}}\frac{1}{\delta r}\sum\limits_{\textrm{S}} \frac{\sum\limits_{i \, \in \,  r\pm\Delta r/2 ; \, P_T^{i,\textrm{track}} \,\in\, \textrm{range}}
P_T^{i,\textrm{track}}}{P_T^{\textrm{jet}}} \, ,	
\end{equation}
where $N_{\textrm{S}}$ is the number of subleading jets entering the analysis and $\delta r$ is the annulus width. This choice of the track momentum allows us to focus on tracks that are soft enough to receive a sizeable contribution from the transverse medium kicks while at the same time are hard enough to survive the strong quenching they experience through the plasma. (Although this argument is only valid at partonic level, the imprint of the described mechanism is also observed in the hadronic distributions displayed in Fig. \ref{fig2}).\

The notable sensitivity of such an observable to the value of the parameter $K$ suggests the possibility to extract the precise strength of the Gaussian broadening mechanism through a comparison with data (using measurements as those in \cite{CMSjettrack}). Ideally, one could also expect measuring deviations from such a Gaussian-like behaviour, indicating rare, but only power-law-rare, hard momentum transfers arising when jet partons scatter off quasi-particles in the soup \cite{DEramo:2012uzl,Kurkela:2014tla}.
\subsection*{Medium response to the jet passage}
\label{back}
The energy and momentum extracted from the jet is assumed to thermalize rapidly into the plasma, as it is most natural in a strongly coupled framework. Within our hybrid model, we have so far considered that the QGP has no memory of the direction of motion of the jet after thermalization, contributing to the uncorrelated background - an assumption which cannot be correct as both energy and momentum have to be conserved. Inevitably, some of this energy carried by the plasma along the jet direction will be reconstructed as part of a jet, leading to non-trivial modifications of the jet properties.\

\begin{figure}[t]
\centering 
\includegraphics[width=.5\textwidth]{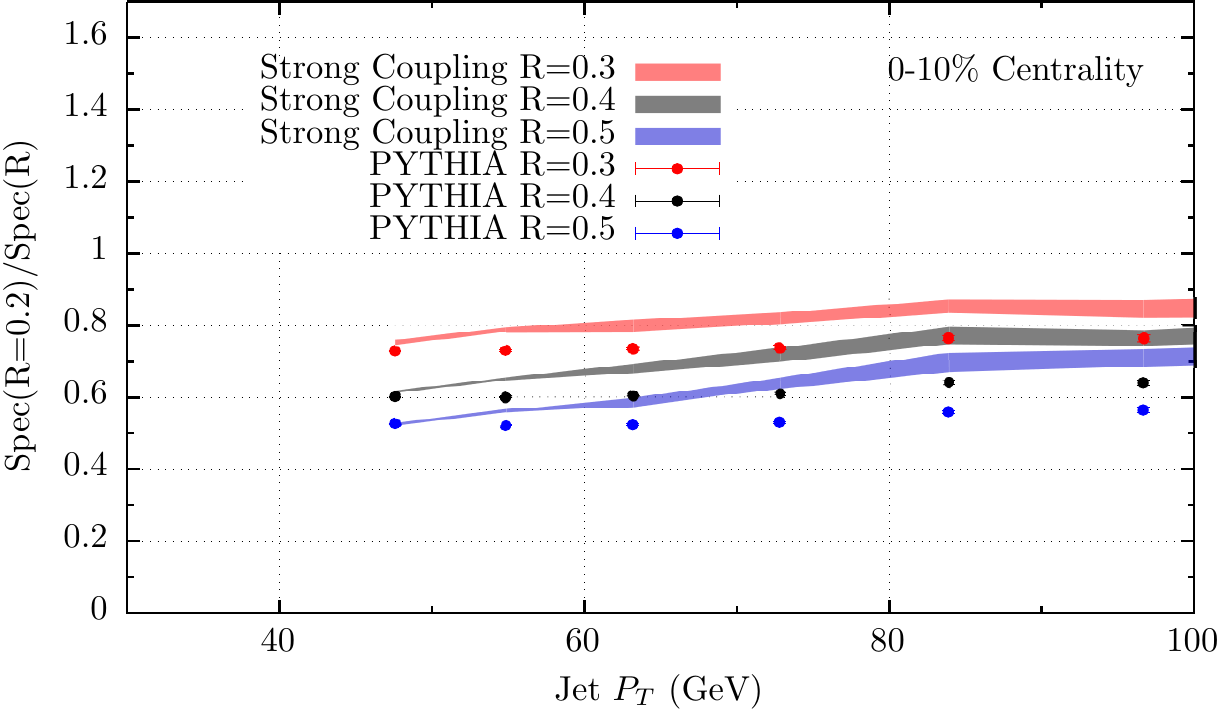}
\caption{Ratios of jet spectra for $R=0.2$ over spectra for $R>0.2$, using PbPb ($0-10\%$ centrality) or pp (PYTHIA) data only. This observable allows for an important cancellation of systematic uncertainties, and can be used to greatly constrain the assumptions on energy redistribution that the different models can have.}
\label{fig3}
\end{figure}

We characterize the medium response to the deposition of momentum from the jet, that is the collective response known as the wake, by studying the induced velocity and temperature variations of the hydrodynamic behaviour of the QGP. By observing that the amount of energy transferred from the jet is around two orders of magnitude below the total energy in the event per unit rapidity, we make the assumption that these variations can be treated as perturbations. For simplicity, we also assume that the unperturbed fluid can be well described by a boost invariant flow, and moreover we consider the perturbation to be narrow in rapidity along the jet direction. With these simplifying working assumptions, we can determine through a Cooper-Frye prescription that the particle spectrum coming from the wake is given by  
\begin{equation}
\begin{split}
& E \frac{dN}{d^3p}=\frac{1}{32 \pi}\frac{m_T}{T^5}\textrm{cosh}(y-y_j)e^{-\frac{m_T}{T}\textrm{cosh}(y-y_j)}\\
&\big[ p_T\Delta P_T \textrm{cos}(\phi-\phi_j)+\frac{1}{3}m_T\Delta M_T\textrm{cosh}(y-y_j) \big],
\end{split}
\end{equation}
where $\Delta P_T$ and $\Delta M_T$ are the lost transverse momentum and mass by the jet respectively, with $y_j$ its rapidity and $\phi_j$ its azimuthal angle  \cite{Casalderrey-Solana:2016jvj}. It is important to note that the last equation has been obtained by assuming production of particles which carry momentum of the order of the temperature, and it is not meant to describe the spectrum of harder particles. We will discuss next the implications on a subset of observables that the introduction of this mechanism has.\

\begin{figure}[t]
\centering 
\includegraphics[width=.5\textwidth]{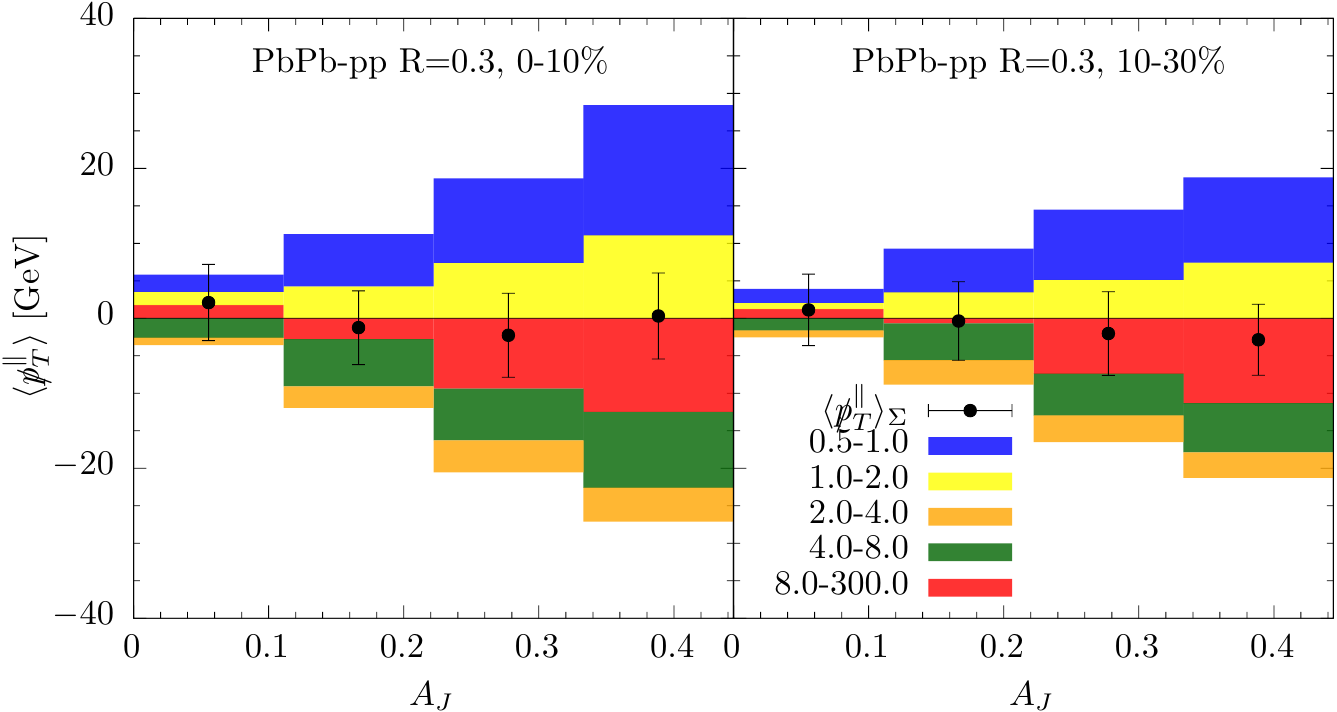}
\caption{Distributions for $\left<\mpt\right>$ selected according to dijet asymmetry $A_J$. What is shown is the difference between PbPb and pp, for two different centralities $0-10\%$ and $10-30\%$. The contributions are binned according to the track $P_T$, showing an excess of soft particles in the subleading (positive) jet side compensated by an excess of hard particles in the leading (negative) jet side. The subleading jet has lost more energy than the leading jet; momentum conservation means that the soft wake produced by the subleading jet must be greater than that produced by the leading jet.}
\label{fig4}
\end{figure}

By studying jet $R_{AA}$ for different jet radius $R$, we have access to  partial information on where the deposited energy has gone to.
In Fig. \ref{fig3} we present an alternative version of such a measurement that consists in taking ratios of either pp or PbPb spectra only, for a fixed radius $R=0.2$ in the numerator and a different, larger radius $R$ in the denominator, where the cancellation of systematic uncertainties is notable. A first conclusion one can draw from the fact that the medium results are always above the vacuum ones is that wider jets have lost more energy than the narrower ones, which can be understood by noting that wider jets contain more partons and therefore more sources of energy loss (as pointed out in other works both from strong coupling \cite{Rajagopal:2016uip} and weak coupling \cite{Milhano:2015mng} points of view). Indeed, such separation grows with jet $P_T$ precisely because the average number of tracks also grows with increasing momentum (and more so the bigger the jet radius $R$). The energy coming from the wake has been transported to larger angles and therefore cannot be recaptured by the relatively narrow jets with $0.2\leq R \leq 0.5$.\

We turn now to the discussion of the so called ``missing-pt'' observables. These characterise the overall energy balance in dijet events by projecting the momentum of all the particles along the dijet axis as quantified by the variable  $\mpt \equiv - P_T \cos\left(\phi_{\textrm{dijet}}-\phi\right)$ (positive for the subleading, negative for the leading), and further binning the contributions as a function of the track $P_T$. In Fig. \ref{fig4} we show one such observable in which events have been classified according to the dijet asymmetry variable $A_J \equiv (P_T^L-P_T^S)/(P_T^L+P_T^S)$. In order to focus only on medium induced modifications, what is shown corresponds to the difference between PbPb and pp, for two different centrality classes. One can see how the more asymmetric a dijet is, the more energy has been transformed into softer particles due to quenching, as can be noted from the excess of the $0.5<P_T<1$ GeV and $1<P_T<2$ GeV bins in the subleading jet side, with less of an excess the more peripheral the collision is. Overall transverse momentum conservation is manifest by observing the cancellation when integrating over all track $P_T$ bins, as shown by $\left<\mpt\right>_{\Sigma}$. While these results are in qualitative agreement with data \cite{Khachatryan:2015lha}, the disagreement in the semi-hard regime of $2<P_T<4$ GeV tracks motivates further improvements of the model.

\subsection*{Conclusions}
\label{conclusions}
Within the context of a hybrid strong/weak coupling model, we have motivated the inclusion of two new physics mechanisms, namely transverse momentum broadening and medium back-reaction, and their effects on a selected set of observables. Broadening is implemented by adding transverse kicks to the propagating partons within the QGP according to a Gaussian distribution with a width controlled by the parameter $K$. Due to the strong quenching, standard jet observables show very little sensitivity to the precise value of $K$, although by focusing on a specific track momentum range we have found an observable that shows a remarkable dependence.\
The effect of medium back-reaction, which follows from energy-momentum conservation, has been estimated by the study of small velocity and temperature variations on top of a boost invariant flow. The wake generated by the momentum deposition from the jet decays into soft particles at large angles, providing a picture which is in overall qualitative agreement with data but fails to reproduce the measured particle spectrum in the semi-hard regime. This could signify the importance of having a less simplified treatment of medium response, or could point to the need of considering other physics mechanisms such as finite resolution effects. For a more detailed discussion of alternative dynamics see \cite{Casalderrey-Solana:2016jvj}.



\nocite{*}
\bibliographystyle{elsarticle-num}
\bibliography{Pablos_D}







\end{document}